%% file: 2023-arxiv-compmix.tex
\documentclass{article}

\usepackage[final]{neurips_2023}

\usepackage[utf8]{inputenc} 
\usepackage[T1]{fontenc}    
\usepackage{hyperref}       
\usepackage{url}            
\usepackage{booktabs}       
\usepackage{amsfonts}       
\usepackage{nicefrac}       
\usepackage{microtype}      
\usepackage{xcolor}         

\usepackage{latexsym}
\usepackage{graphicx}
\graphicspath{{./images/}}
\usepackage{booktabs} 
\usepackage{color}  
\usepackage{amsmath}  
\usepackage{subcaption}
\usepackage{caption}
\usepackage{tikz}
\usepackage{colortbl} 
\usepackage{framed}
\usepackage{multirow}
\usepackage{multicol}
\usepackage{hyperref}
\usepackage{url}
\usepackage{balance}
\usepackage{verbatim}
\usepackage{cancel}
\usepackage{xspace} 
\usepackage[ruled,vlined]{algorithm2e}
\usepackage{bbold} 
\usepackage{balance}
\usepackage{pifont} 

\newcommand{\struct}[1]{\texttt{\small #1}}
\newcommand{\utterance}[1]{\textit{#1}}
\newcommand{\phrase}[1]{\textit{``#1''}}

\newenvironment{Snugshade}[1][236,236,236]{
    \setlength{\itemsep}{0pt}
     \setlength{\parsep}{0pt}
     \setlength{\topsep}{0pt}
     \setlength{\partopsep}{0pt}
     \setlength{\leftmargin}{1.5em}
     \setlength{\labelwidth}{0em}
     \setlength{\labelsep}{0em} 
    \setlength{\parskip}{0pt}
    \definecolor{shadecolor}{RGB}{#1}%
    \begin{snugshade}
}{%
    \end{snugshade}%
}

\newcommand{\convinse}{\textsc{Convinse}\xspace}
\newcommand{\convmix}{\textsc{ConvMix}\xspace}
\newcommand{\compmix}{\textsc{CompMix}\xspace}
\newcommand{\explaignn}{\textsc{Explaignn}\xspace}

\newcommand{\clocq}{\textsc{Clocq}\xspace}

\newcommand{\unikqa}{\textsc{UniK-QA}\xspace}

\newcommand{\gpt}{\textsc{Gpt-3}\xspace}

\newcommand{\cmark}{\ding{51}}%
\newcommand{\xmark}{\ding{55}}%

\makeatother

\begin{document}


\title{CompMix: A Benchmark for Heterogeneous\\Question Answering}

%

\author{%
  Philipp Christmann, Rishiraj Saha Roy, and Gerhard Weikum\\
  Max Planck Institute for Informatics\\
  Saarland Informatics Campus\\
  Saarbruecken, Germany \\
  \texttt{\{pchristm, rishiraj, weikum\}@mpi-inf.mpg.de}
}


\newcommand{\squishlist}{
    \begin{list}{$\bullet$}{ 
        \setlength{\itemsep}{0pt}
        \setlength{\parsep}{1pt}
        \setlength{\topsep}{1pt}
        \setlength{\partopsep}{0pt}
        \setlength{\leftmargin}{1.5em}
        \setlength{\labelwidth}{1em}
        \setlength{\labelsep}{0.5em} 
    } 
}

\newcommand{\squishend}{
  \end{list}  }
  
\newcommand{\GW}[1]{{\color{blue}{GW: #1}} }
\newcommand{\PC}[1]{{\color{orange}{PC: #1}} }
\newcommand{\RR}[1]{{\color{red}{RR: #1}} }

\newcommand{\myparagraph}[1]{\noindent \textbf{#1}.}

\setcounter{secnumdepth}{4}


\maketitle
\input{sections/00-abstract}
\input{sections/01-introduction}
\input{sections/04-benchmark}
\input{sections/05-experiments}
\input{sections/07-formalities}
\input{sections/08-conclusion}


\balance

\bibliographystyle{ACM-Reference-Format}
\bibliography{compmix}

\end{document}

%% file: sections/00-abstract.tex
\begin{abstract}
Fact-centric question answering (QA)
often requires access to multiple, heterogeneous, information sources.
By jointly considering several sources like a knowledge base (KB), a text collection, and tables from the web,
QA systems can enhance their answer coverage and confidence.
However, existing QA benchmarks are mostly constructed with
a single 
source
of knowledge 
in mind.
This limits
capabilities
of these benchmarks
to fairly evaluate QA systems that 
can tap into more than one information repository.
To bridge this gap, we release \compmix, a 
crowdsourced
QA benchmark which naturally demands
the integration of
a mixture of input sources.
\compmix has a total of $9{,}410$ questions,
and features several complex intents like joins and temporal conditions.
Evaluation of a range of QA systems 
on \compmix
highlights
the need for further research on leveraging
information from heterogeneous sources.
\end{abstract}


%% file: sections/01-introduction.tex
\section{Introduction}
\label{sec:intro}

\myparagraph{Motivation} 
The goal in
factual
question answering (QA) is to derive crisp
answers to information needs issued by end users~\cite{saharoy2022question}.
There has been a long line of
research on
fact-based
QA, that can largely be divided into
three main directions:
(i)
methods
that use a large curated knowledge base (KB)
like Wikidata~\cite{vrandevcic2014wikidata},
YAGO~\cite{suchanek2007yago} or DBpedia~\cite{auer2007dbpedia} as information source (KB-QA)~\cite{abujabal2018never, bast2015more, bhutani2019learning, vakulenko2019message},
(ii)
systems
that
retrieve
information from a text corpus (text-QA)~\cite{izacard2021leveraging, chen2017reading, zhang2021answering},
and 
(iii)
works
that
answer
questions based on a set of web tables (table-QA)~\cite{jauhar2016tables, chakrabarti2020open, herzig2021open}.
Each of these directions has its own benchmarks
that are frequently used for developing, testing and comparing QA systems~\cite{dubey2019lc, talmor2018web, berant2013semantic, bordes2015large, yang2018hotpotqa, herzig2021open, kwiatkowski2019natural, yih2016value}.

However, using only a single information source limits the \textit{answer coverage} of QA systems:
the individual sources are not complete, and may fail to cover the 
knowledge
required for answering a user question.
Consider, as an example, the question below:
\begin{Snugshade}
    \utterance{Who was fouled before the first penalty in the 2022
    FIFA
    final?}
\end{Snugshade}
This kind of
detailed
information
on a sports event
is rarely covered in a structured information source like a KB or table,
but can be found in text discussing the content of the match.
On the other hand, structured sources
often
include information that is
not present in text.
Tables often store match-specific details, and would contain, for instance, the answer to the following question:
\begin{Snugshade}
    \utterance{Argentina's ball possession in the 2022 WC final?}
\end{Snugshade}
For some questions, 
answers appear 
in multiple sources.
Such \textit{answer redundancy} can also be helpful for QA systems, and
boost their \textit{confidence} in predicted answers. For instance, consider:
\begin{Snugshade}
    \utterance{In which stadium was the 2022
    soccer
    world cup final played?}
\end{Snugshade}
The answer to this question 
occurs
in a Wikipedia infobox, text content, 
and Wikidata.
It may even be necessary to join evidence from multiple sources
for answering a more complex question:
\begin{Snugshade}
    \utterance{Which team was behind by two goals but still won a FIFA final?}
\end{Snugshade}
The list of FIFA World Cup finals and their winners could be looked up in a KB,
but the goal deficit information associated with the match timeline would either be discussed in text, or 
could be reasoned over statistics in tables. 
These observations have triggered
work on heterogeneous QA~\cite{sun2018open,sun2019pullnet,oguz2021unikqa,savenkov2016knowledge,xu2016question,xu2016hybrid, xiong2019improving}:
%
jointly harnessing multiple sources
for answering
factual
questions~\cite{saharoy2022question}.

\myparagraph{Limitations of state-of-the-art}
There are currently three strategies of evaluating heterogeneous QA: (i) using benchmarks for single-source QA but showing that using more sources improves performance~\cite{xu2016hybrid,xu2016question,savenkov2016knowledge,oguz2021unikqa}; (ii) using benchmarks for single-source QA, but artificially removing parts of the ``main'' source before augmenting the benchmark with new sources~\cite{sun2018open,sun2019pullnet}; and (iii) using dedicated benchmarks for heterogeneous QA~\cite{talmor2018web,chen2020hybridqa}.
The first approach usually leads to quick saturation on benchmarks: all answers are still available only in the primary source, which is what the methods primarily target, and auxiliary sources bring in incremental gains. 
The second approach is inherently flawed because considering heterogeneous sources obviously improves performance, as the main source is intentionally weakened.
This creates an artificial situation 
and does not expose the true strengths and weaknesses of methods built for heterogeneous QA.


Our contribution belongs to the third approach.
There are a few existing benchmarks for multi-source QA~\cite{talmor2018web,miller2016key,zhang2018variational},
but these either contain synthetic
questions 
and do not reflect idiosyncrasies in formulation and intent concerning real users,
or cover only a narrow spectrum of sources and domains~\cite{chen2021finqa,chen2020hybridqa,zhu2021tat,li2021tsqa,chen2021open}.

\myparagraph{A new benchmark} 
We make the case for a benchmark that
\textit{inherently requires} the
usage of
a mixture of
information sources,
as a more natural testbed for evaluating heterogeneous QA systems.
To this end, we release \compmix
(\underline{Comp}lete questions over a \underline{Mix}ture of sources),
a \textit{crowdsourced} QA benchmark with questions that require heterogeneous sources for answering
(Wikidata KB, and Wikipedia text, tables and infoboxes).
The dataset has $9{,}410$ questions created by \textit{humans} from \textit{five different domains}: 
books, movies, music, TV series and soccer.
The answers are grounded to the Wikidata KB,
which
allows
use of 
consistent evaluation
metrics
for 
QA systems
returning either entity IDs or simple strings.

\myparagraph{Contributions}
This paper
presents
our benchmark \compmix,
accompanied by
an in-depth analysis. 
We identify complex phenomena
in the questions,
like
temporal
conditions,
multiple entities and relations,
aggregations
and
comparisons.
We investigate the effect of combining multiple sources on answer coverage
and redundancy,
and 
show that heterogeneous sources are 
truly
required.

Finally, we evaluate multiple recent heterogeneous QA methods on
\compmix,
and identify questions for which none of these systems 
gives correct answers.
Interestingly, the results for a recent GPT model
show that 
even a large language model (LLM) can answer only half of the questions for this realistic and challenging benchmark.
The \compmix benchmark is publicly available at \textbf{\url{https://qa.mpi-inf.mpg.de/compmix}}.

%% file: sections/04-benchmark.tex
\section{Benchmark description}
\label{sec:compmix}

\begin{table} [t] \small
	\centering
	\caption{Comparing benchmarks for heterogeneous QA.}
    \newcolumntype{H}{>{\setbox0=\hbox\bgroup}c<{\egroup}@{}}
	\resizebox*{\columnwidth}{!}{
    \begin{tabular}{l | c c c c H | c c c}
        \toprule
            \textbf{Dataset}                &       \textbf{KB}       & \textbf{Text}  & \textbf{Table} & \textbf{Info}  & \textbf{Images} & \textbf{OR}         &  \textbf{HQ}  & \textbf{OD}   \\
        \midrule
            \textsc{HybridQA}~\cite{chen2020hybridqa}       & \xmark     &  \cmark            &   \cmark   & \xmark  &   \xmark  & \xmark   & \cmark & \cmark   \\
            
            \textsc{MultiModalQA}~\cite{talmor2021multimodalqa}   & \xmark     &  \cmark            &   \cmark   & \xmark  &   \cmark  & \cmark & \xmark  & \cmark     \\
            
            \textsc{OTT-QA}~\cite{chen2021open}         & \xmark     &  \cmark            &   \cmark   & \xmark  &   \xmark   &   \cmark  & \cmark   & \cmark  \\
            
            \textsc{ManyModalQA}~\cite{hannan2020manymodalqa}                & \xmark     & \cmark   & \cmark  & \xmark  &   \cmark   & \xmark & \cmark & \cmark         \\ 

            
            \textsc{WikiMovies}~\cite{miller2016key}                & \cmark     & \cmark   & \xmark  & \xmark &   \xmark  & \cmark & \xmark   & \xmark      \\        
            

            \textsc{TAT-QA}~\cite{zhu2021tat}                & \xmark     & \cmark   & \cmark  & \xmark  &   \xmark  & \xmark & \cmark   & \xmark      \\        

            \textsc{FinQA}~\cite{chen2021finqa}                & \xmark     & \cmark   & \cmark  & \xmark &   \xmark  & \xmark & \cmark   & \xmark      \\        

            \textsc{HetPQA}~\cite{shen2022product}                & \xmark     & \cmark   & \cmark  & \cmark &   \xmark  & \xmark & \cmark   & \xmark      \\        
        \midrule
            \compmix (ours)               & \cmark     & \cmark   & \cmark  & \cmark &   \xmark  & \cmark & \cmark   & \cmark      \\            
        \bottomrule

    \end{tabular}
    
    }	
    \\ \textbf{OR:} Open Retrieval; \textbf{HQ:} Human Questions; \textbf{OD:} Open Domain.
	\label{tab:benchmarks}
\end{table}

\begin{table} [t] \small
	\centering
	\caption{Basic statistics for the \compmix benchmark.}
        \begin{tabular}{p{5cm} p{7cm}}
            \toprule
            Domains                 & Books, Movies, Music, TV series, Soccer           \\
            Questions               & $9{,}410$ (train: $4{,}966$, dev: $1{,}680$, test: $2{,}764$) \\
            \midrule
            Avg. question length         & $9.19$ words  (min=$2$, median=$9$, max=$28$) \\
            Avg. no. of question entities     & $1.11$ (min=$1$, median=$1$, max=$4$)\\
            Avg. answer length (text) & $2.17$ words (min=$1$, median=$2$, max=$21$) \\
            Avg. no. of answers             & $1.02$ (min=$1$, median=$1$, max=$6$) \\ 
            Entities covered        & $5{,}413$ (long-tail: $2{,}511$, with <$50$ KB-facts) \\
            \bottomrule
        \end{tabular}
	\label{tab:statistics}
\end{table}



\subsection{Prior benchmarks and \compmix rationale} 

There are many datasets for KB-QA (like WebQuestions~\cite{berant2013semantic}, SimpleQuestions~\cite{bordes2015large}, and CSQA~\cite{saha2018complex}),
text-QA (like SQuAD~\cite{rajpurkar2016squad}, HotpotQA~\cite{yang2018hotpotqa}, and NaturalQuestions~\cite{kwiatkowski2019natural}),
and table-QA (like WikiTableQuestions~\cite{pasupat2015compositional}, NQ-Tables~\cite{herzig2021open}, and WikiSQL~\cite{zhong2017seq2sql}).
However, these benchmarks were created with
the intention of having
a specific underlying
source for answering, 
which already contains almost all answers to the questions.
This
restricts their utility 
as a testbed for heterogeneous QA.
Thus, existing work on heterogeneous QA,
being forced to rely on these benchmarks, 
would often remove significant chunks of information from this ``main'' information source ($\simeq 50\%$ of Freebase removed for evaluating on WebQuestions in~\cite{sun2019pullnet}),
and add parts of other sources to
simulate a 
setting
with heterogeneous sources.

All existing benchmarks for heterogeneous QA suffer from one or more of the following issues:
(i) their questions are not fully human-generated, and hence lack the 
diverse formulations
of real users~\cite{talmor2018web,zhang2018variational,miller2016key};
(ii) they are restricted to small or artificial KBs, orders of magnitude smaller than large curated knowledge bases like Wikidata or DBpedia~\cite{miller2016key,zhang2018variational};
(iii) they span only two sources, like tables and text~\cite{chen2021finqa,zhu2021tat,chen2021open}, or text and knowledge bases~\cite{talmor2018web,miller2016key,zhang2018variational}; 
(iv) they explore only one 
domain like finance~\cite{chen2021finqa,zhu2021tat}, geography~\cite{li2021tsqa}, or e-commerce~\cite{shen2022product}; and
(v) their questions are only in 
conversational form with implicit intent, unsuitable for evaluating stand-alone QA methods~\cite{christmann2022conversational,deng2022pacific,nakamura2022hybridialogue}.

\compmix removes these shortcomings: 
(i) it is crowdsourced; (ii) it includes the full KB as one of the knowledge sources; (iii) it spans four sources; (iv) it covers five domains; and (v) it contains self-contained complete questions. A succinct comparison
of salient properties across benchmarks
is in Table~\ref{tab:benchmarks}. 

\subsection{\compmix} 

\begin{figure}[t]
    \begin{center}
         \includegraphics[width=0.7\columnwidth]{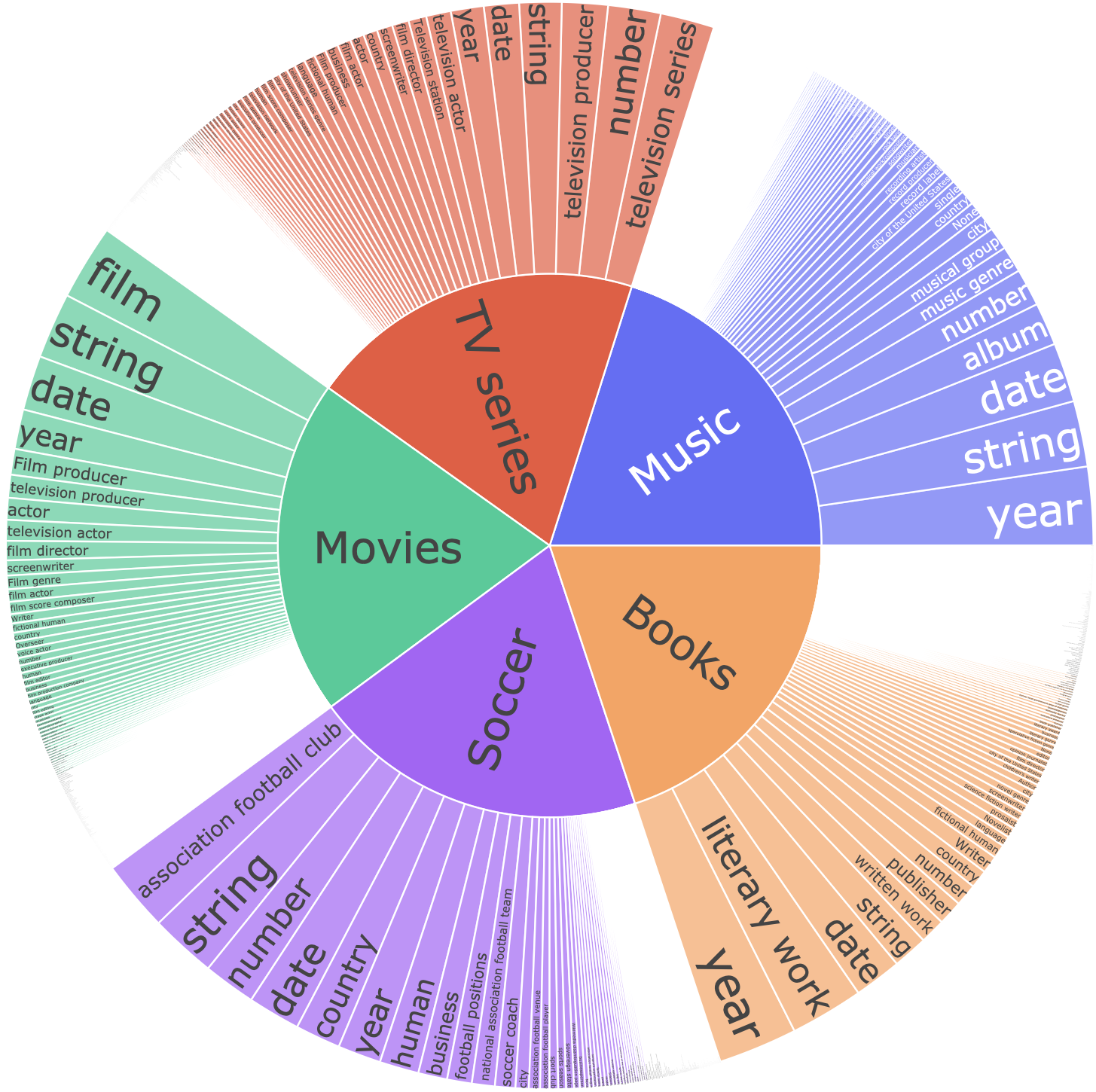}
         \caption{Answer-type frequencies per domain in \compmix.}
         \label{fig:answer-types}
    \end{center}
\end{figure}

We create \compmix by collating the \textit{completed} (intent-explicit) versions of the
potentially incomplete (intent-implicit) questions in the \convmix~\cite{christmann2022conversational} benchmark,
which is a dataset for conversational QA over heterogeneous sources.
These completed questions are provided directly by crowdworkers on Amazon Mechanical Turk (AMT),
i.e. are created by \textit{humans}.
The answers to the questions were derived from \textit{four} sources: either the full Wikidata KB,
or the text, tables or infoboxes from all of Wikipedia.
The questions span \textit{five} domains: movies, tv series, music, books, and soccer (a distribution of expected answer types for each domain is in Fig.~\ref{fig:answer-types}).
Overall, the benchmark comprises $9{,}410$ questions,
split into train set ($4{,}966$),
development set ($1{,}680$),
and test set ($2{,}764$).
Basic statistics for \compmix can be found in Table~\ref{tab:statistics}.
A notable property of our dataset is the presence of a significant fraction of questions with \textit{long-tail entities} (last row), a major vulnerability of LLM methods.

\compmix includes questions, their domains, and their corresponding answers. Answers are 
Wikidata 
entity identifiers (text labels are also provided), plaintext strings, or normalized dates.
This enables consistent evaluation across extractive and generative answering models.
In addition, 
entity markup in question formulations are provided by crowdworkers.
Answer sources are given, too:
\phrase{KB}, \phrase{text}, \phrase{table}, or \phrase{infobox}.


\section{Benchmark analysis}
\label{sec:analysis}

\subsection{Answer coverage}

One key desideratum of the benchmark is that heterogeneous sources are actually required for answering the questions inside.
To verify that this is the case, we analyzed the answer coverage of each information source, which is the number of questions that a source contains the answer for.
In a good benchmark for heterogeneous QA, each source should have an answer coverage far less than $100\%$.

At the time of benchmark creation, Turkers were given a domain, and they picked up an entity of choice from the domain, followed by asking a natural question using this entity, and then provided an answer to the question.
They also provided the source they consulted for locating their answer.
For computing coverage, we first consider these \textit{source annotations by the crowdworkers}.
However,
this measurement
only captures whether a specific information source has the desired information,
without any implications concerning the other sources.

Therefore, we also conducted an automatic analysis of the answer coverage
using a
recall-oriented
retriever that, given a question, tries to obtain as many relevant pieces of evidence
as possible
from all our sources. 
This retriever is implemented as in~\cite{christmann2022conversational,christmann2023explainable},
and would first disambiguate
KB-entities from the question (using \clocq~\cite{christmann2022beyond}, a recent system),
and then retrieve KB-facts, text-sentences, table-records and infobox-entries with these disambiguated KB-entities.
For each evidence,
mentions of entities are
linked
to the KB.
We measure this \textit{automated
answer coverage}
as the number of questions for which the gold answer
is among this set of mentioned entities in the pool of retrieved evidence. As with any large-scale automated analysis, this statistic is a noisy proxy, because the mere presence of an answer does not necessarily mean that the surrounding evidence is question-relevant.


The results of both analyses are in Table~\ref{tab:answer-presence}.
First, we see that the AMT annotators used the KB, text and infoboxes almost equally often to answer their questions (tables also consulted $\geq\!\!10\%$ of times).
This proves that \compmix is \textit{not biased} towards any specific underlying source.
Second, from the automated measurement, we learn that adding an information source always improves the answer coverage.
Note that this is a natural expansion, as opposed to augmentation after artificial suppression of large parts of specific sources.
By including all sources, the answer coverage goes up to about $87\%$. 
Note that our recall-oriented retriever only provides a loose upper bound: the performance of an actual retriever that balances recall and precision would currently reach a lower number (cf. Sec.~\ref{sec:experiments}).
Thus, our benchmark leaves substantial room for the development of smart \textit{heterogeneous retrievers}.
Overall, these measurements suggest that all four sources are naturally required for answering the questions in \compmix,
and different sources complement each other nicely.

\subsection{Answer redundancy}

Answer redundancy creates scope to test a heterogeneous system's ability to boost confidence in its prediction when multiple matches happen across sources.
For each question, we thus measured the number of sources touched by the retrieved pieces of evidence that actually contain the gold answer.
Results are in Table~\ref{tab:redundancy}.
What we can see from here is that for a substantial proportion of questions, the answer is located in two  ($\simeq 17\%$)  or three  ($\simeq 34\%$), out of four, sources.
A reasonable chunk even has redundancy across all sources ($\simeq 20\%$).
This shows that \compmix has ample answer redundancy to be exploited by some appropriate heterogeneous QA model.

\subsection{Anecdotal examples}

For each of our five domains, Table~\ref{tab:compmix} shows representative examples from the \compmix benchmark.
The examples illustrate that our dataset has a wide range of questions in terms of both \textit{syntactic structure} -- from well-formulated fluent questions
(1, 4, 5, 9)
to ad hoc telegraphic queries
(6, 7),
as well as \textit{semantic complexity} -- from simple intents
(6, 8)
to 
more complex ones
requiring conjunction (2),
temporal understanding (3, 5),
or aggregations (9).



\begin{table} [t] 
    \begin{center}
        \caption{Answer coverage across information sources.} 
        \newcolumntype{G}{>{\columncolor [gray] {0.90}}c}
        \newcolumntype{H}{>{\setbox0=\hbox\bgroup}c<{\egroup}@{}}
        	\begin{tabular}{l G c} 
            \toprule
                \textbf{Source(s)} & \textbf{Annotated}  & \textbf{Automated}  \\
                \midrule
                    \textbf{KB}             &  $0.308$  &  $0.807$  \\
                    \textbf{Text}           &  $0.280$  &  $0.690$  \\
                    \textbf{Tables}         &  $0.112$  &  $0.272$  \\
                    \textbf{Infoboxes}           &  $0.299$  &  $0.545$ \\
                \midrule
                    \textbf{KB+Text}        &  $0.588$  &  $0.853$   \\
                    \textbf{KB+Tables}      &  $0.420$  &  $0.821$   \\
                    \textbf{KB+Infoboxes}        &  $0.607$  &  $0.831$   \\
                    \textbf{Text+Tables}    &  $0.393$  &  $0.702$  \\
                    \textbf{Text+Infoboxes}      &  $0.580$  &  $0.734$  \\
                    \textbf{Tables+Infoboxes}    &  $0.412$  &  $0.610$  \\
                \midrule
                    \textbf{KB+Text+Tables}        &  $0.701$  &  $0.857$   \\
                    \textbf{KB+Text+Infoboxes}      &  $0.888$  &  $0.861$   \\
                    \textbf{KB+Tables+Infoboxes}        &  $0.720$  &  $0.841$   \\
                    \textbf{Text+Tables+Infoboxes}    &  $0.692$  &  $0.743$  \\
                \midrule
                    \textbf{All sources}    &  $1.000$  &  $0.865$ \\
                \bottomrule
        \end{tabular}
        \label{tab:answer-presence}
    \end{center}
\end{table}

\begin{table} [t] 
    \begin{center}
        \caption{Answer redundancy across information sources.}
        \newcolumntype{G}{>{\columncolor [gray] {0.90}}c}
        \newcolumntype{H}{>{\setbox0=\hbox\bgroup}c<{\egroup}@{}}
        	\begin{tabular}{l G G} 
            \toprule
                Answer found in \textbf{$\boldsymbol{1}$ source}          &  $0.157$  \\
                Answer found in \textbf{$\boldsymbol{2}$ sources}         &  $0.168$  \\
                Answer found in \textbf{$\boldsymbol{3}$ sources}         &  $0.341$  \\
                Answer found in \textbf{\textbf{all} sources}             &  $0.199$  \\
            \bottomrule
        \end{tabular}
        \label{tab:redundancy}
    \end{center}
\end{table}


\begin{table*} [t]
	\centering 
	\caption{Representative questions from \compmix. Sources that can be used for answering these questions are in brackets.}
	\resizebox*{\columnwidth}{!}{
	   \begin{tabular}{p{3cm} | p{3cm} | p{3cm} | p{3cm} | p{3cm}}
           \toprule
			     \textbf{Books}	&	\textbf{Movies}		&	\textbf{Music}	&	\textbf{TV series} &	\textbf{Soccer} \\
           \midrule
			     1. \utterance{What did Rayford Steele from Left Behind do as a job?}	&	2. \utterance{Which lead actress appeared in both Terms of Endearment and The Evening Star?} &	3. \utterance{Who replaced Ozzy Osbourne in Black Sabbath the first time?}	&	4. \utterance{What TV show featured the character called Carrie Mathison?}		&	5. \utterance{Where did the Uruguay national football team play their first recorded match?}  \\
			     \struct{Pilot}	&	\struct{Shirley MacLaine}	&	\struct{Ronnie James Dio}	&	\struct{Homelande}	&	\struct{Paso del Molino} \\ \relax
			     [KB, Text] 	&	[KB]	&		[Text, Info] 	&		[KB, Text, Info] 	&	 [Text] 	\\
           \midrule
			     6. \utterance{Author of the book To Kill a Mockingbird?}	&	7. \utterance{Film in which Wallace Reid played the role of Walter Jarvis?}	&	8. \utterance{What is the singer Lemmy's birth name? }	&	9. \utterance{How many episodes of The 100 did Jason Rothenberg write?} &	10. \utterance{Who was runner up in the 1998 World Cup?} \\
			     \struct{Harper Lee}	&	\struct{The Ghost Breaker}	&	\struct{Ian Fraser Kilmister}	&	\struct{16}	&	\struct{Brazil football team} \\ \relax
			      [KB, Text, Table, Info] 	&	[KB, Text]	&		[KB, Text, Info] 	&		[KB, Text, Table] 	&		[KB, Text, Info] \\
            \midrule
			     11. \utterance{Name the fifth book in Malory Towers series.}	&	12. \utterance{Which movie is longer, Hamlet or Gone with the Wind?}		&	13. \utterance{What year was Inna's Hot album released in the US? }	&	14.  \utterance{Which season of Teen Wolf did Tyler Posey become a co-producer?} &	15. \utterance{Which soccer player scored the most number of goals in the UEFA Euro 2004 tournament?} \\
			     \struct{In the Fifth at Malory Towers}	&	\struct{Hamlet}	&	\struct{2009}	&	\struct{5}	&	\struct{Milan Baroš} \\ \relax
			     [KB, Table] 	&	[KB, Info]	&		[Text] 	&		 [Text, Info]	&		[KB, Text, Info, Table]\\
            \midrule
			     16. \utterance{What years were the two volumes of Little Women published?}	&	17. \utterance{What is the run time of Titanic?}	&	18. \utterance{What is the name of the second single in the album Arise?}	&	19. \utterance{What year was Matt Groening born?}	&	20. \utterance{Who was the kit manufacturer of Chelsea Football Club from 1981 to 1983?} \\
			     \struct{1868, 1869}	&	\struct{195 minutes}	&	\struct{Dead Embryonic Cells}	&	\struct{1954}	&	\struct{Le Coq sportif} \\ \relax
			     [KB, Info] 	&	[KB, Infobox]	&		[Text, Table] 	&		[Text] 	&		[Text, Table] \\
	       \bottomrule
	   \end{tabular}
    }
	\label{tab:compmix}
\end{table*}

%% file: sections/05-experiments.tex
\section{Evaluation with \compmix}
\label{sec:experiments}

\myparagraph{Metrics} 
We use standard QA metrics for evaluating models on \compmix:
(i) \textit{Precision at 1} ({P@1}), which is either 1 or 0 according as the top-ranked system answer is correct or not;
(ii) \textit{Mean reciprocal rank} ({MRR}), which is the reciprocal of the first rank at which a correct answer is located;
and, (iii) \textit{Hit at 5} ({Hit@5}), which is either 1 or 0 according as the first five system responses contains a gold answer or not.
A system answer is considered correct if it exactly (case-insensitive) matches a Wikidata
ID
(if QA system returns IDs)
or the accompanying plaintext string/entity label
(if QA system returns simple text).
Metrics are averaged over all questions.

\myparagraph{Models}
To better understand the state-of-the-art in heterogeneous QA,
we evaluate several recent QA models
that incorporate heterogeneous sources
on \compmix. We also include GPT in our model suite, to verify if LLMs trained on colossal web corpora are already sufficient for this task.
We compare the following models:
\squishlist
    \item \textbf{\unikqa~\cite{oguz2021unikqa}} follows a retriever-reader pipeline,
and verbalizes evidence from each source
into text.
DPR~\cite{karpukhin2020dense}
retrieves
relevant evidences
from the verbalized text, and a Fusion-in-decoder (FiD) model~\cite{izacard2021leveraging}
generates the answer.
Due to unavailability of end-to-end source code,
we approximate \unikqa
by replacing DPR with BM25~\cite{robertson2009probabilistic}.
FiD generates strings, that are mapped to a ranked list of KB items, by following~\cite{christmann2023explainable}.
    \item \textbf{\convinse~\cite{christmann2022conversational}} is method
    for conversational QA over heterogeneous sources, but can also be applied to complete questions.
    It derives an intent-explicit structured representation for a question, and feeds this into a retriever-reader pipeline.
    \item \textbf{\explaignn~\cite{christmann2023explainable}} 
    is another method for heterogeneous QA that makes use of iterative graph neural networks for deriving the answer instead of a generative reader model like FiD.
    \item \myparagraph{\gpt}
For evaluating \gpt~\cite{brown2020language} (model: \textit{text-davinci-003}), we use the following prompt, which performed the best among different alternatives:
\phrase{Please answer the following question by providing the crisp answer entity, date, year, or numeric number. Q: <question>}.
The generated answer string is then compared with the label and KB-aliases of the gold answer(s), to allow for potential synonymy (all strings lowercased).
P@1 = 1 for exact matches, and zero otherwise.
\gpt generates only a single answer, and thus metrics for ranked lists are inapplicable.
\squishend

\myparagraph{Results} Findings in Table~\ref{tab:main-res}
reveal two key takeaways:
(i) systems from the literature only reach about $45\%$ P@1 on \compmix, showing substantial room for model improvement.
Much higher numbers have been reported for the compared
models in previous sub-optimal evaluation settings
(\unikqa reaches $80\%$ accuracy on WebQuestionsSP):
this highlights challenges in \compmix;
(ii) The task is far from solved for 
LLMs, with the P@1 reached by \gpt being merely $50\%$.
We attribute this to 
a large number of rare
and emerging entities in our benchmark (see Table~\ref{tab:statistics}).
To put aggregate performance in perspective, we found that for $2{,}764$ questions
($81.9$\%),
at least one of the methods 
\textit{failed} to produce a correct answer.
On the other hand, for $759$
($27.5$\%)
\textit{none} of the methods (including \gpt) could find the correct answer.
Table~\ref{tab:anec} shows one such unanswered question per domain.
The second and fifth question make a perfect case for merging multiple sources, as subtle cues like \phrase{adult Pi Patel} or
\phrase{twin brothers} are likely to be mentioned in textual sources, while movie cast or club membership is more easily looked up via structured repositories.


\begin{table} [t]
\begin{center}
    \caption{Heterogeneous QA models on \compmix (test set).}
    \newcolumntype{G}{>{\columncolor [gray] {0.90}}c}
    \begin{tabular}{l G G G}
        \toprule
            \textbf{Method $\downarrow$ / Metric $\rightarrow$} & \textbf{P@1}  & \textbf{MRR}  & \textbf{Hit@5} \\ 
        \midrule
            \textbf{\unikqa~\cite{oguz2021unikqa}}
            & $0.440$ &   $0.467$ & $0.494$  \\
        

            \textbf{\convinse~\cite{christmann2022conversational}}
            & $0.407$  &   $0.437$ & $0.483$   \\
    
            \textbf{\explaignn~\cite{christmann2023explainable}} 
            & $0.442$ &   $0.518$ & $0.617$  \\

            \textbf{\gpt~\cite{brown2020language}}  (text-davinci-003)
            & $0.502$ &   $-$ & $-$  \\
        \bottomrule
    \end{tabular} 
    \label{tab:main-res}
\end{center}
\end{table}

\begin{table} \small
    \caption{Anecdotal questions for which none of the tested methods could derive the correct answer.}
    \begin{tabular}{p{8cm}}
        \toprule
            \utterance{What was the original title of the book Twilight? } \\
            \utterance{Who played as adult Pi Patel in Life of Pi movie?} \\
            \utterance{What album is the song Closing Time on?} \\
            \utterance{Who composed the theme music for the TV series Fury? } \\ 
            \utterance{Who were the twin brothers who played soccer for Manchester United?} \\
        \bottomrule
    \end{tabular} 
	\label{tab:anec}
\end{table}

%% file: sections/07-formalities.tex
\section{Data Sharing and Ethics}
\label{sec:formalities}

\myparagraph{Licensing}
The \compmix benchmark is licensed under a
Creative Commons Attribution 4.0 International License\footnote{\url{http://creativecommons.org/licenses/by/4.0/}}.

\myparagraph{Availability}
The benchmark is 
released
on our project website\footnote{\url{https://qa.mpi-inf.mpg.de/compmix/}},
with inclusion of a leaderboard to keep track of the state-of-the-art.
\compmix is also offered at Hugging Face 
for a broader audience\footnote{\url{https://huggingface.co/datasets/pchristm/CompMix}}.
The DOI of \compmix is \textbf{\url{https://doi.org/10.57967/hf/0707}}.

\myparagraph{Ethical considerations}
\compmix collates completed questions from the \convmix benchmark.
For collecting \convmix, human annotators from AMT asked factoid questions in a conversational setting.
No personal or other critical data was collected or published.
The \compmix benchmark does not contain any personal or other critical data.
All questions are provided anonymously.
The annotators for collecting the \convmix dataset were paid
a fair compensation for their work,
consistent with the German minimum wage
(irrespective of their residential country).

%% file: sections/08-conclusion.tex

\section{Conclusion}
\label{sec:conclusion}

We release \compmix, a benchmark for heterogeneous QA that inherently requires the usage of multiple sources.
Answering questions in \compmix requires systems to work consistently well for intents spread across five domains, and deal with a wide variety of challenging human formulations asking about rare entities.
Thus, our hope is that this resource can help facilitate progress in
developing more robust QA models that can appropriately exploit complementary
and potentially redundant sources of information.
A promising direction for improvement would be to include questions that need answers of a different flavor of heterogeneity:
sentences, passages, or longer lists.
